\documentclass[preprint,review,11pt]{elsarticle}

\usepackage{lineno,hyperref}
\usepackage{amsthm}
\usepackage{amsmath,amsfonts,amssymb}
\usepackage{tikz}
\usepackage{tikz-3dplot}
\usepackage{prettyref}
\usepackage{subfigure}
\usepackage{float}
\usepackage{booktabs}
\usepackage[margin=3.2cm]{geometry}

\hypersetup{colorlinks=true, linkcolor=blue, citecolor=blue, urlcolor=blue}
\newrefformat{fig}{Fig.~\ref{#1}}
\newrefformat{tbl}{Table~\ref{#1}}
\newrefformat{sec}{Section~\ref{#1}}
\newrefformat{eq}{Eq.~\ref{#1}}
\newrefformat{alg}{Algorithm?~\ref{#1}}

\makeatletter
\def\ps@pprintTitle{%
   \let\@oddhead\@empty
   \let\@evenhead\@empty
   \def\@oddfoot{\reset@font\hfil\thepage\hfil}
   \let\@evenfoot\@oddfoot
}
\makeatother

\biboptions{authoryear}

\begin{document}


\begin{frontmatter}

\title{The dynamic energy balance in earthquakes expressed by fault surface morphology}

\author[cumt]{Xin Wang\corref{cor1}}
\ead{ericrussell@zju.edu.cn}
\author[cumt]{Juan Liu}
\author[cumt]{Feng Gao}
\author[cumt]{Zhizhen Zhang}
\cortext[cor1]{Corresponding author}

\address[cumt]{State Key Laboratory for Geomechanics and Deep Underground Engineering, China University of Mining and Technology, Xuzhou 221116, China}


\begin{abstract}

The dynamic energy balance is essential for earthquake studies. The energy balance approach is one of the most famous developments in fracture mechanics. To interpret seismological data, crack models and sliding on a frictional surface (fault) models are widely used. The macroscopically observable energy budget and the microscopic processes can be related through the fracture energy $G_c$. The fault surface morphology is the direct result of the microscopic processes near the
crack tip or on the frictional interface. Here we show that the dynamic energy balance in
earthquakes can be expressed by fault surface morphology, and that they are quantitatively
linked. The direct shear experiments proves the predictions of the theoretical discussions, and show that the strain rate has crucial influence on the dynamic energy balance.

\end{abstract}


\end{frontmatter}


\section{Introduction}

An earthquake may be considered to be a dynamically running shear crack \citep{Scholz2019}, in which the dynamic energy balance is essential for earthquake studies. In the field of fracture mechanics, the energy balance approach employed by Griffith has become one of the most famous developments in materials science \citep{Collins1993}. To interpret seismological data, crack models are often used in part because the theories on cracks have been developed well. On the other hand, seismic faulting may be more intuitively viewed as sliding on a frictional surface (fault) where the physics of friction, especially stick slip, plays a key role.

The fracture energy, $G_c$, in crack theory is the energy needed to create new crack surfaces near the crack tip. Thus, the system must expend the threshold fracture energy $G_c$ before the crack can extend. In contrast, in frictional sliding model, $D_c$, is introduced as a critical slip before rapid sliding begins at a constant friction. If the initial stress $\sigma_0$ of the system drops more or less linearly to the final stress $\sigma_1$, i.e., the final value of the frictional stress $\sigma_f$, the energy spent in the system before this happens can be approximately written as $\frac{1}{2} (\sigma_0 - \sigma_1)D_c$. Thus, if we are to link a crack model to a friction model, we can equate this energy to $G_c$, i.e., $G_c = \frac{1}{2} (\sigma_0 - \sigma_1)D_c$. Then we can relate the macroscopically observable energy budget to the microscopic processes in a surprisingly general way. Any constraint on fracture energy obtained from the energy budget will provide a strong bound on all microscopic rupture processes \citep{Kanamori2004}. \citet{Svetlizky2014} have shown that interface rupture propagation is described by, essentially, the same framework of the crack model. This suggests that an analogous 'Griffith' condition may exist for the onset of rupture propagation for a frictional interface.

The fault surface morphology is the direct result of the microscopic processes near the crack tip or on the frictional interface. Here we show that the dynamic energy balance in earthquakes can be expressed by fault surface morphology, and that they are quantitatively linked.

\section{The description of fault surface morphology and its links to the dynamic energy balance in earthquakes}

The description of fault surface morphology has been investigated in the literature \citep[e.g.,][]{Ladanyi1969,Barton1977,Plesha1987,Saeb1992,Amadei1998} while trying to study the contribution of surface morphology to the shear strength of rock fractures. As shearing strictly depends on three-dimensional contact area location and distribution \citep{Gentier2000}, \citet{Grasselli2002} proposed a method for the quantitative three-dimensional description of a rough fracture surface, and based on this description, \citet{Grasselli2003} proposed a constitutive criterion to model the shear strength of fractures. \citet{Wang2019} defined the terms ``quasi steps'' and ``quasi striations'' to refer to morphological structures that are created during the creation of new crack surfaces and the friction on the frictional interface. The same morphological structures are also causing anisotropy of the fracture surface morphology, hence the anisotropy of the surface morphology's contribution on its shear strength. The terms ``quasi steps'' and ``quasi striations'' are broader definitions of the fault steps and fault striations whether they can be obviously seen or not. The parameter $\theta^*_{max}/C$ was proposed by \citet{Grasselli2002} to describe the contribution of fracture surface morphology on the shear strength. \citet{Wang2019} proposed a theoretical model that describes the contribution of quasi steps and quasi striations on the shear strength. And by fitting this theoretical model to $\theta^*_{max}/C$ from each slip direction on the fracture surface, the amount of quasi steps $R_G$ and quasi striations $R_H$ can be estimated. Combined with the method proposed by \citet{Wang2017} for outcrop fracture surface extraction from point cloud data, the estimated quasi steps and quasi striations data had good applications in tectonics \citep[e.g.,][]{Wang2020}.

\begin{figure}[H]
  \centering
  \includegraphics[width=0.4\textwidth]{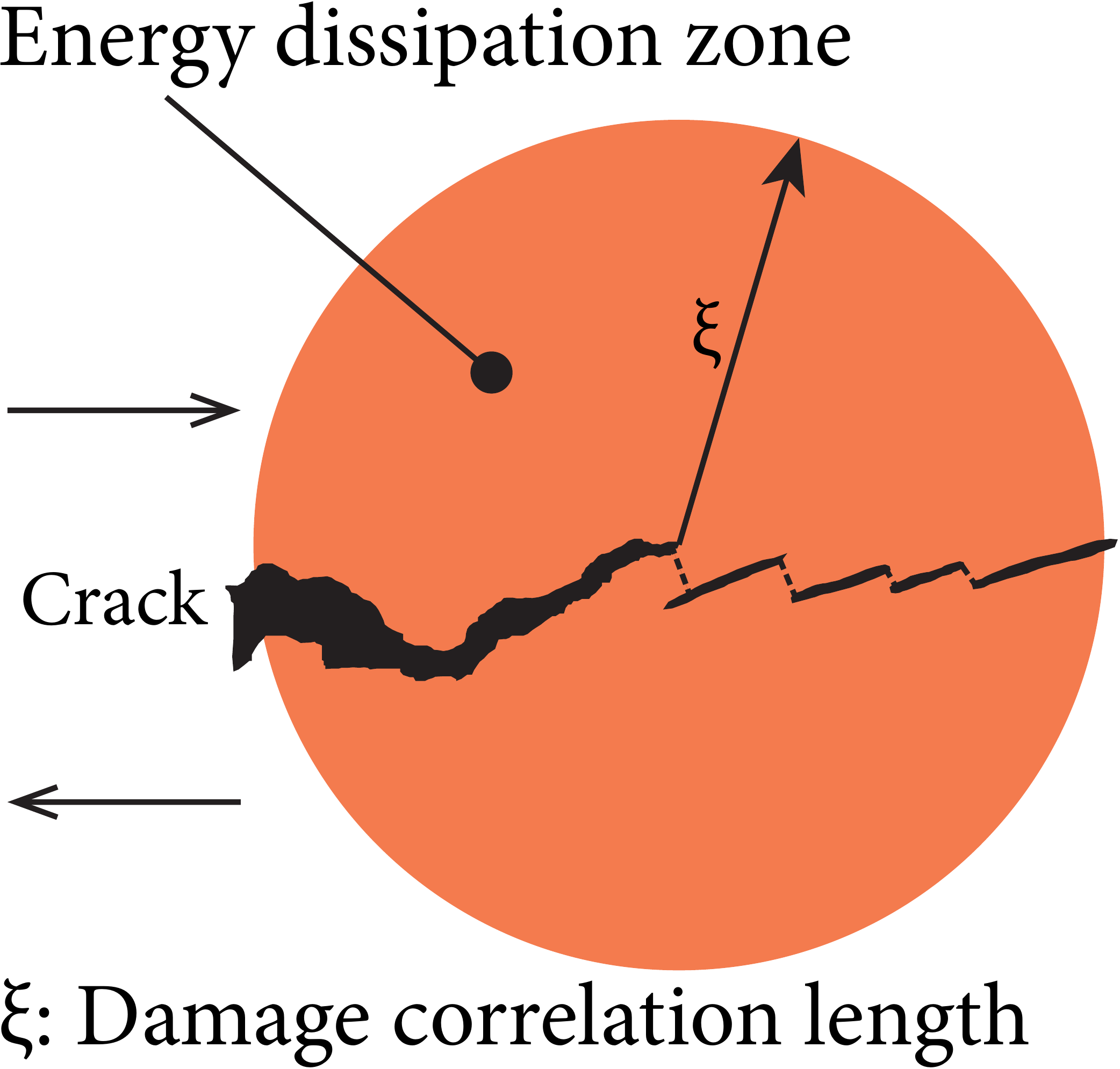}
  \caption{The model of the tip growth of mode II (in-plane shear) cracks. The formation of voids are documented in \citet{Bouchaud1999}.}
  \label{fig:crack_model-out}
\end{figure}

Let's consider the formation of quasi steps during the crack growth. The voids (\prettyref{fig:crack_model-out} as suggested by \citet{Bouchaud1999}) are nucleated under the influence of the stress field adjacent to the tip, but not at the tip, due to the existence of the plastic zone that cuts off the purely linear-elastic (unphysical) crack-tip singularities. The crack grows by coalescing the voids with the tip, creating a new stress field which induces the nucleation of new voids \citep{Bouchbinder2004}. The morphological structures of quasi steps are then formed during the crack growth under shear load as iillustrated in \prettyref{fig:crack_model-out}. A crucial aspect of this picture is the existence of a typical scale, $\xi_c$, which is roughly the distance between the crack tip and the first void, at the time of the nucleation of the latter. In this picture there exists a ‘‘energy dissipation zone
’’ in front of the crack tip in which plastic yield is accompanied by the evolution of damage cavities. From this picture, it can also be seen that the typical scale $\xi_c$ is positively related with the estimated amount of quasi steps $R_G$ since the quasi steps are more developed on the fracture surface with larger value of $\xi_c$.

A simple model for $\xi_c$ was developed \citep{Bouchbinder2004,Afek2005} by assuming the energy dissipation zone to be properly described by the Huber–von Mises plasticity theory. The material yields as the distortional energy exceeds a material-dependent threshold $\sigma_Y^2$ and the typical distance $\xi_c$ scales as \citep{Bouchbinder2004}
\begin{equation}
 \xi_c \sim \frac{K_{II}^2}{\sigma_Y^2},
 \label{eq:scale_xi_c}
\end{equation}
where $K_{II}$ is the stress intensity factor for mode II (in-plane shear) cracks. On the other hand, the linear-elastic solution is still valid outside the energy dissipation zone, and the energy release rate $G^*$ can be expressed as
\begin{equation}
 G^* = \frac{K_{II}^2}{E'},
\end{equation}
where $E'$ is related to Young's modulus $E$ and Poisson's ratio $\nu$ (for plane strain):
\[E' = \frac{E}{1 - \nu^2}.\]
The definition of the energy release rate $G^*$ is
\begin{equation}
 G^* = \frac{\partial \Omega}{\partial s},
\end{equation}
where $\Omega$ is the strain energy and $s$ is the crack growth area. Then \prettyref{eq:scale_xi_c} can be rewritten as
\begin{equation}
 \xi_c \sim G^*/\frac{\sigma^2_Y}{E'} = \frac{\partial \Omega}{\partial s}/\frac{\sigma^2_Y}{E'}.
\end{equation}
The amount of quasi steps $R_G$ is a description of the degree of quasi steps development over the fracture surface, so it is an average quantity over the fracture surface. $R_G$ should scale with $\overline{\xi_c}$, the average of $\xi_c$ over the fracture surface $S$:
\begin{equation}
 R_G \sim \overline{\xi_c} = \frac{1}{S}\int_s \xi_c \sim \frac{\Omega}{\gamma_G},
 \label{eq:scale_strain_energy}
\end{equation}
where
\begin{equation}
 \gamma_G = \int\limits_{S} \frac{\sigma^2_Y}{E'}.
 \label{eq:scale_material_property}
\end{equation}

Let's consider the formation of quasi striations during the friction on the frictional interface. As suggested by \citet{Svetlizky2014}, the interface rupture friction is described by, essentially, the same framework of the crack model. The main difference is that the crack model consider the crack growth in the rock mass, while the friction model consider the crack growth through the local protruding obstacles on the frictional interface, which are then removed or deformed during the friction. As the slip goes on, more protruding obstacles that face the slip direction with high angles are removed or deformed while protruding obstacles that are high angle perpendicular to the slip direction are kept. This anisotropy results in the macroscopic quasi striations structures. The amount of quasi striations $R_H$ is a description of the degree of quasi striations development over the fracture surface, so it is an average quantity over the fracture surface. $R_H$ should scale with $E_f$, the frictional energy like this:
\begin{equation}
 R_H \sim \frac{E_f}{\overline{\omega}},
 \label{eq:scale_frictional_energy}
\end{equation}
where $\overline{\omega}$ is the average strain energy needed to remove or deform a protruding obstacle. Suppose $n$ protruding obstacles are removed or deformed, according to \prettyref{eq:scale_strain_energy} and \prettyref{eq:scale_material_property}, the strain energy scale as
\[\omega_1 \sim \int\limits_{S^*_1} \frac{\sigma^2_Y}{E'}, \omega_2 \sim \int\limits_{S^*_2} \frac{\sigma^2_Y}{E'}, ..., \omega_n \sim \int\limits_{S^*_n} \frac{\sigma^2_Y}{E'},\]
where $S^*_i$ is the crack area in the $i$-th protruding obstacle. The average strain energy $\overline{\omega}$ can be written as
\begin{equation}
 \overline{\omega} = \frac{1}{n}\sum^n \omega_i \sim \int\limits_{\overline{S^*}} \overline{\biggl(\frac{\sigma^2_Y}{E'}\biggr)},
\end{equation}
where
\[\overline{S^*} = \frac{1}{n}\sum^n S^*_i,\]
and
\[\overline{\biggl(\frac{\sigma^2_Y}{E'}\biggr)} = \sum^n \int\limits_{S^*_i} \frac{\sigma^2_Y}{E'} \bigg/ \sum^n S^*_i.\]

Finally, we have the link between the fault surface morphology and the dynamic energy balance in earthquakes:
\begin{equation}
 \frac{R_G}{R_H} \sim \frac{\overline{\omega}}{\gamma_G} \frac{\Omega}{E_f}.
 \label{eq:scale_link_morphology_energy}
\end{equation}

\section{The experimental results and discussions}

\begin{figure}[H]
  \centering
  \includegraphics[width=0.8\textwidth]{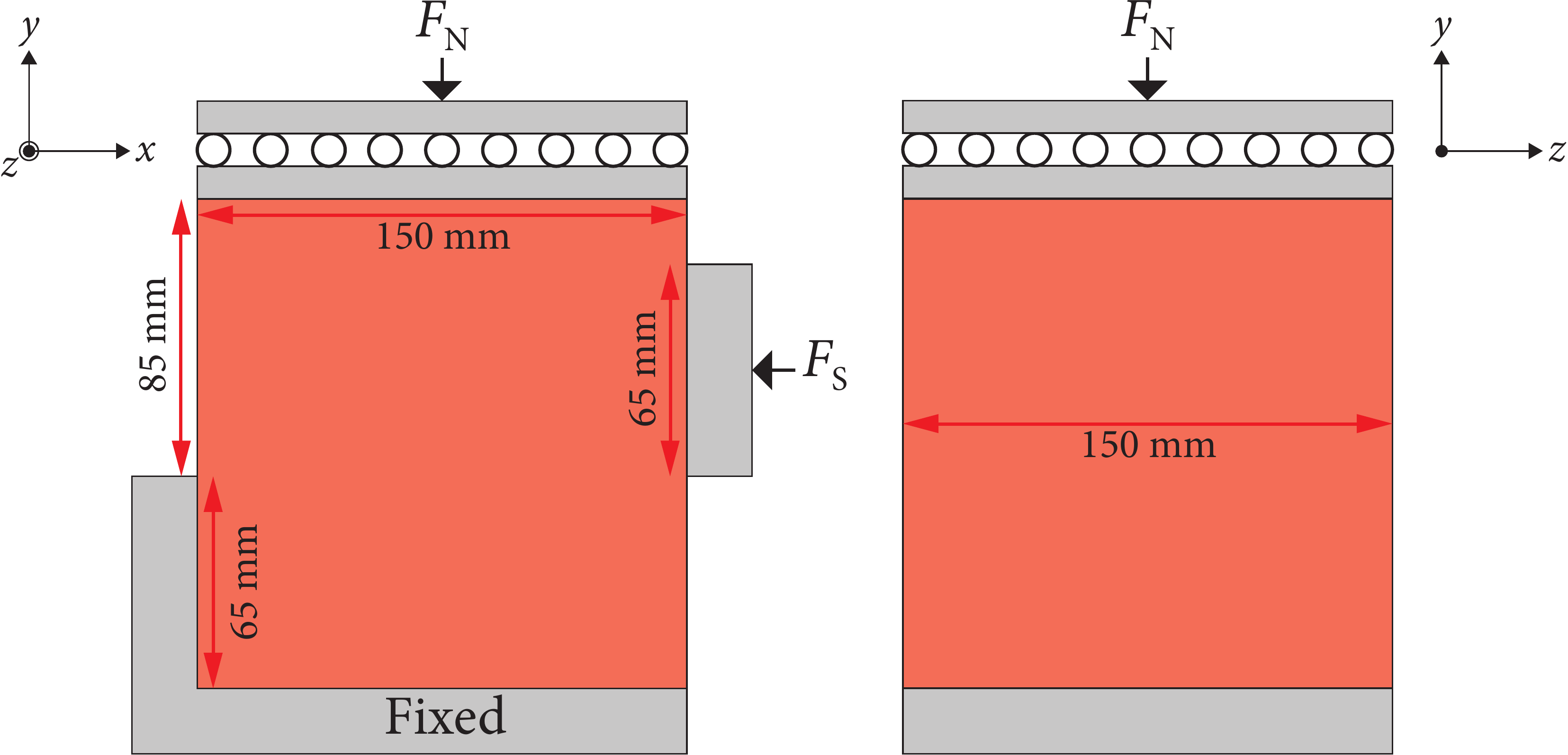}
  \caption{Experiment settings of direct shear test.}
  \label{fig:experiment_settings-out}
\end{figure}

\begin{figure}[H]
  \centering
  \includegraphics[width=0.3\textwidth]{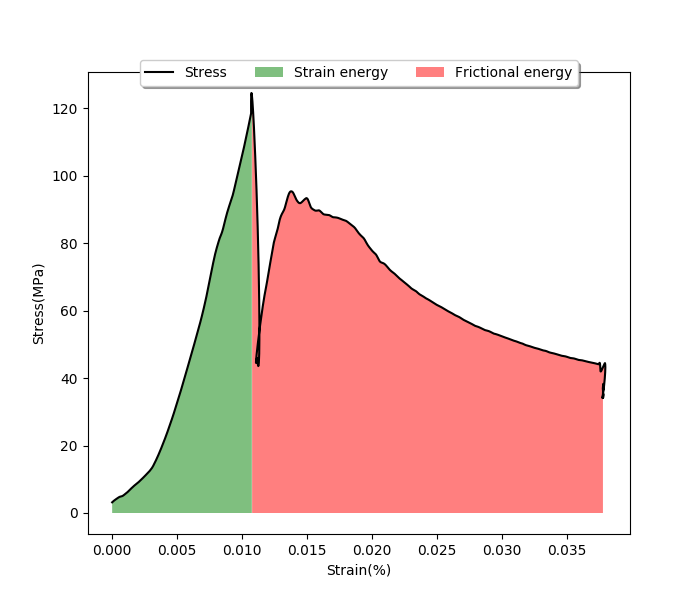}
  \includegraphics[width=0.3\textwidth]{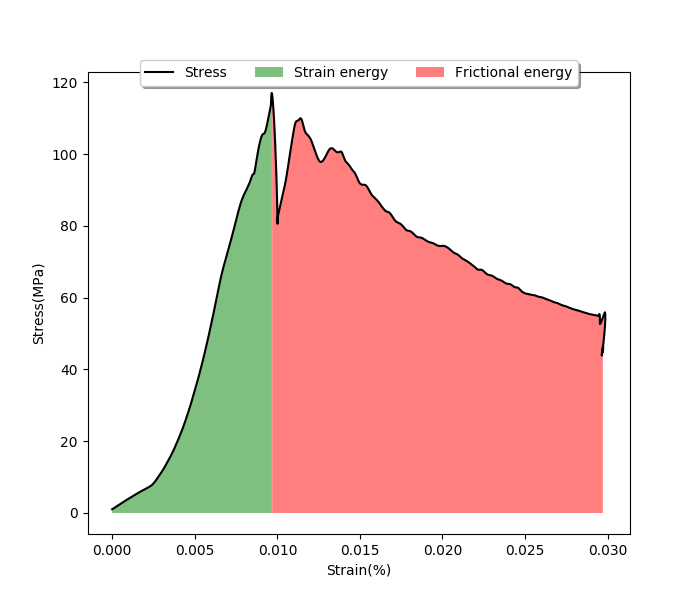}
  \includegraphics[width=0.3\textwidth]{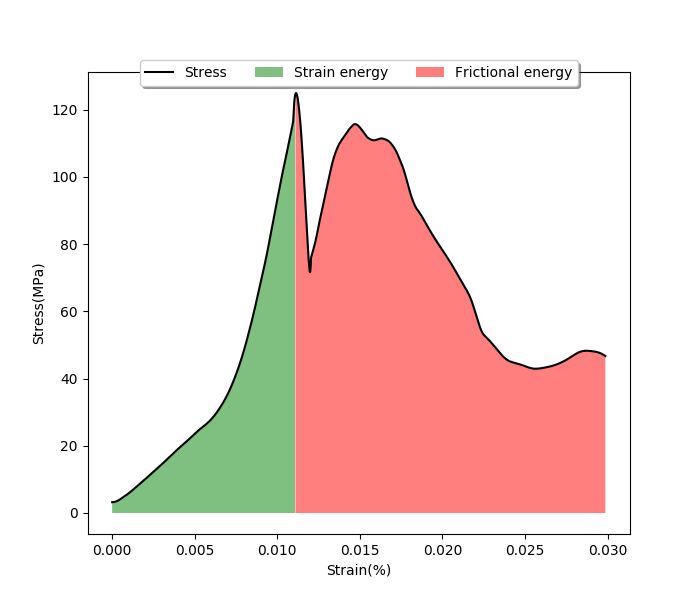}
  \includegraphics[width=0.3\textwidth]{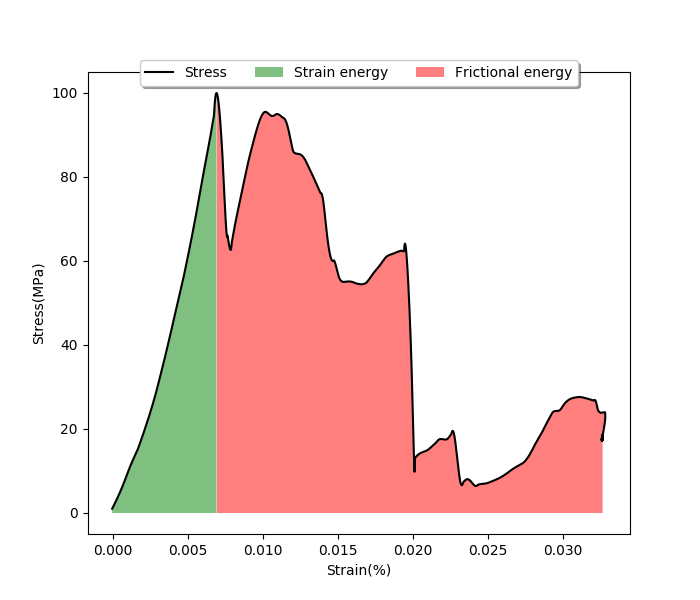}
  \includegraphics[width=0.3\textwidth]{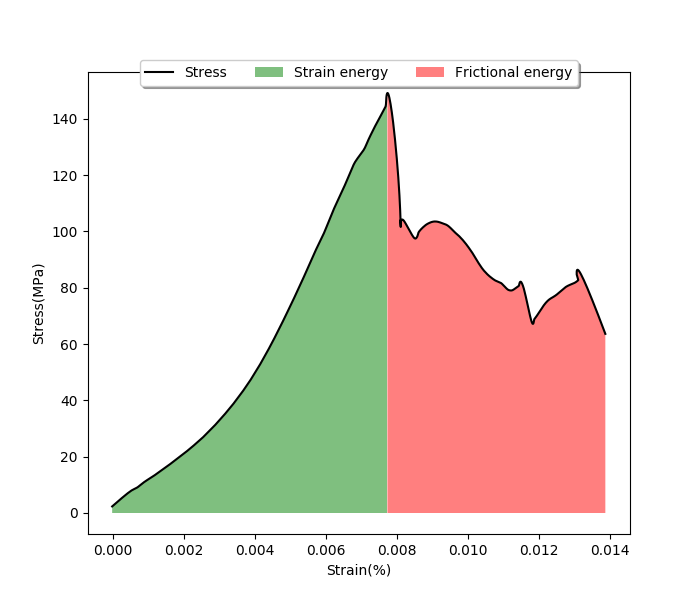}
  \includegraphics[width=0.3\textwidth]{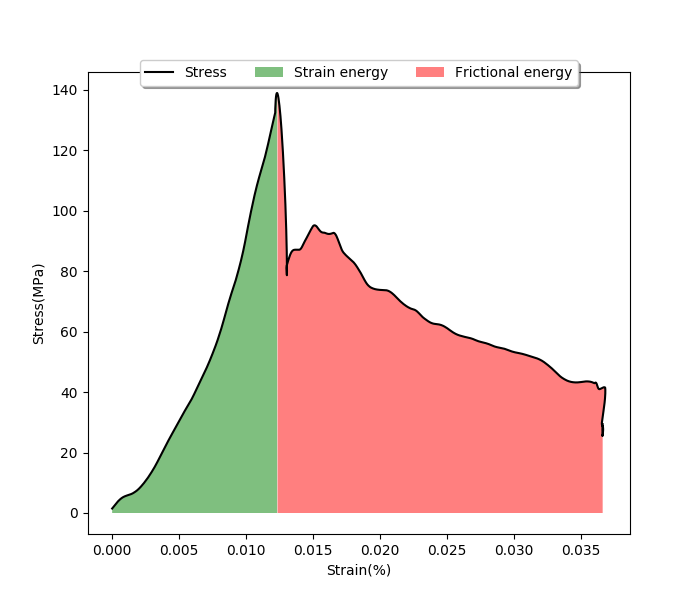}
  \includegraphics[width=0.3\textwidth]{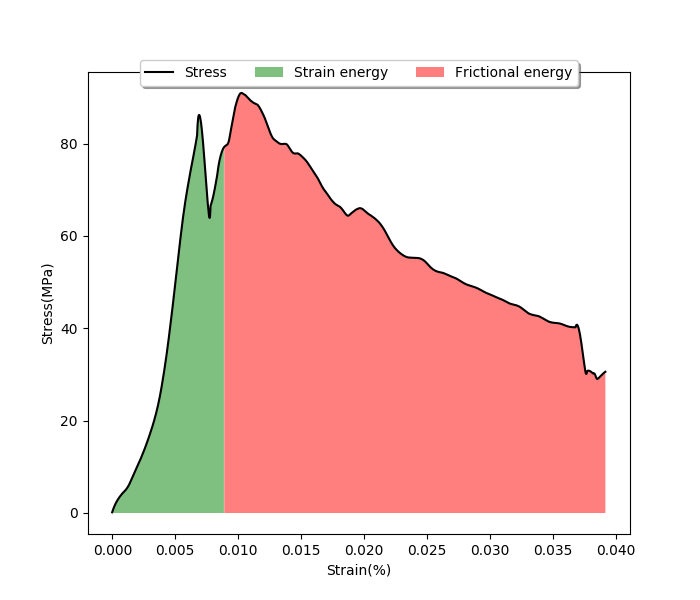}
  \includegraphics[width=0.3\textwidth]{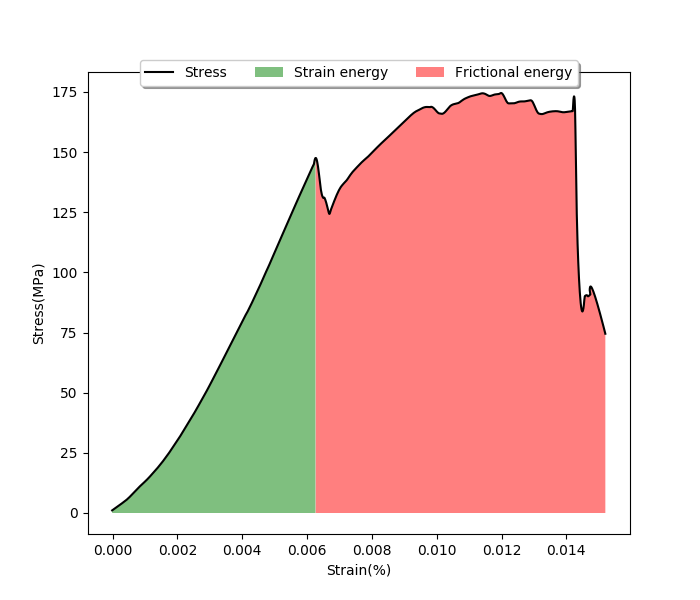}
  \includegraphics[width=0.3\textwidth]{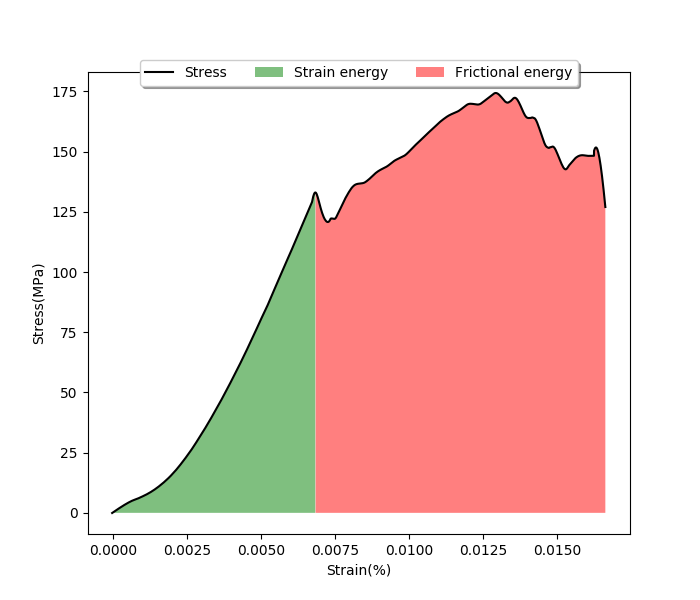}
  \caption{Examples of the stress-strain curve. Roughly the green regions represent the strain energy to be released during the crack growth, and the red regions represent the frictional energy spent during the friction on the frictional interface.}
  \label{fig:strain_energy_frictional_energy}
\end{figure}

An experiment was designed to test the above theoretical discussion. The dimensions and experiment settings are illustrated in \prettyref{fig:experiment_settings-out}. 14 red sandstone rock samples are tested in this simple direct shear experiment with the strain rate ranging from 0.01/min to 0.083/min. The final fracture surfaces after the crack growth and the interface friction processes are scanned and analyzed, the amount of quasi steps $R_G$ and quasi striations $R_H$ are estimated. On the other hand, the stress-strain curves are recorded during the tests. As shown in \prettyref{fig:strain_energy_frictional_energy}, the strain energy is mainly accumulated before the formation of the macroscopic crack through the rock sample, while the frictional energy is mainly spent after the formation of the macroscopic crack. Thus the strain energy $\Omega$ and the frictional energy $E_f$ can be roughly estimated.

\begin{figure}[H]
  \centering
  \includegraphics[width=0.6\textwidth]{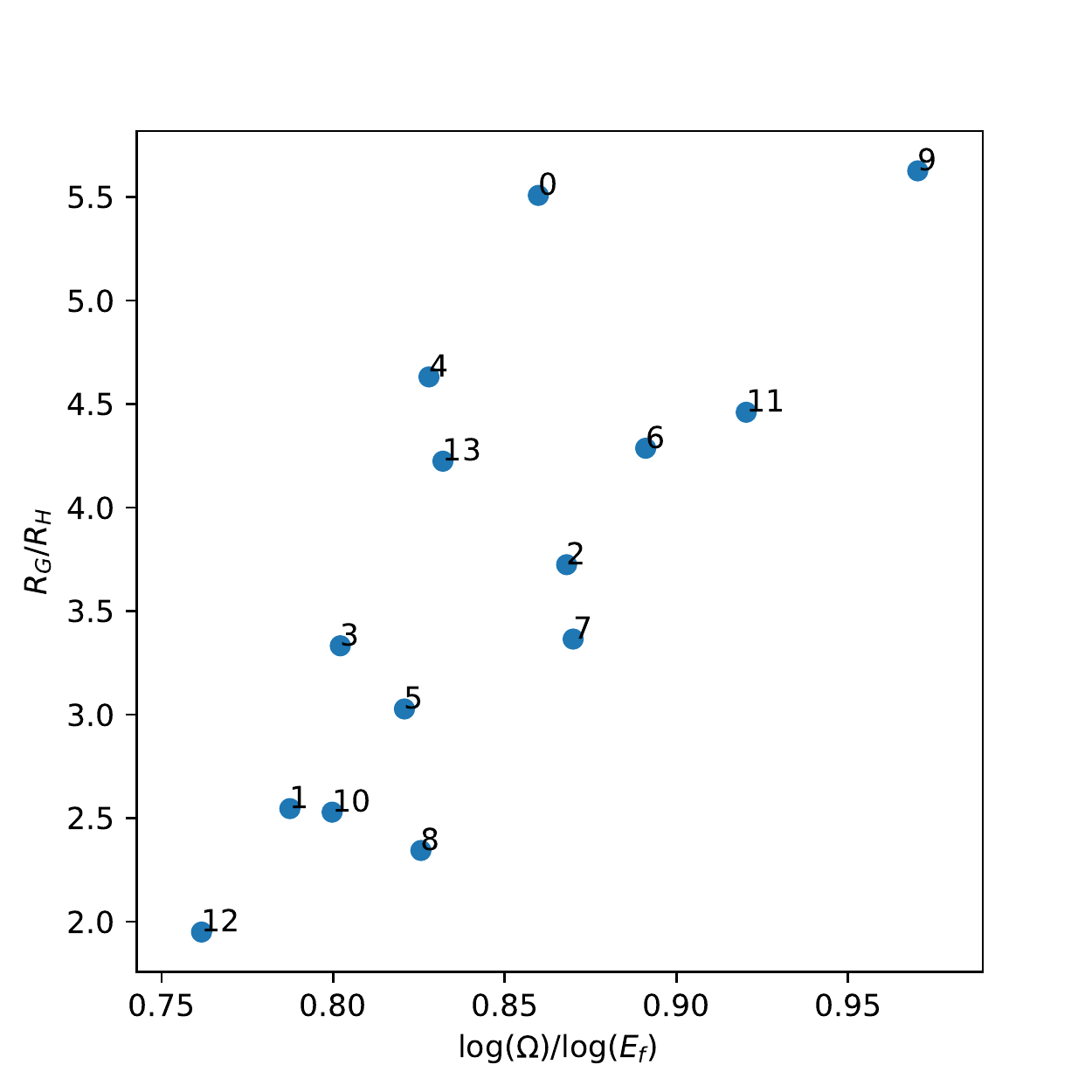}
  \caption{The experimental results. The ratio between the amount of quasi steps and the amount of quasi striations ($R_G/R_H$) is plotted against $\log{(\Omega)}/\log{(E_f)}$. Each labeled point is the result of the corresponding rock sample, and its strain rate can be found in \prettyref{tbl:sample_number_strain_rate}.}
  \label{fig:experimental_results}
\end{figure}

The rock samples are numbered and their strain rate during the test are shown in \prettyref{tbl:sample_number_strain_rate}. The ratio between the amount of quasi steps and the amount of quasi striations ($R_G/R_H$) is plotted against $\log{(\Omega)}/\log{(E_f)}$. The logarithm of the strain energy and the frictional energy are taken here in order to compare their values with $R_G$ and $R_H$, although the logarithm relationship is not explicitly shown in above discussions (e.g., \prettyref{eq:scale_strain_energy} and \prettyref{eq:scale_frictional_energy}). Results of rock samples are scatter ploted in \prettyref{fig:experimental_results} with their sample numbers labeled on their corresponding scatter points for reference.

\begin{table}[H]
   \centering
   \caption{Rock sample numbers and their strain rate during the test.} 
   \begin{tabular}{{c|ccccccc}}
      \toprule
      Sample N.O. & 0 & 1 & 2 & 3 & 4 & 5 & 6 \\
      \midrule
      Strain rate (\%/min)& 0.08 & 0.077 & 0.07 & 0.063 & 0.037 & 0.03 & 0.01 \\
      \midrule
      Sample N.O. & 7 & 8 & 9 & 10 & 11 & 12 & 13 \\
      \midrule
      Strain rate (\%/min) & 0.017 & 0.023 & 0.033 & 0.043 & 0.057 & 0.067 & 0.083 \\
      \bottomrule
   \end{tabular}
   \label{tbl:sample_number_strain_rate}
\end{table}

The results show that there seems indeed a quantitative link between the fault surface morphology and the dynamic energy balance in earthquakes as theoretically predicted above. Some of the scattered results show a constant value of $\overline{\omega}/{\gamma_G}$ (e.g., sample number 12, 1, 10, 5, 2, 6, 11 and 9) because $\overline{\omega}$ and $\gamma_G$ are both integrals of $\sigma^2_Y/E'$ over some area, whose average tend to be the same for the same rock sample, although it varies locally. Thus the link predicted by \prettyref{eq:scale_link_morphology_energy} has the property of material-independent, at least for dry brittle materials discussed here.

Note that sample number 3, 13, 4 and 0 seem have slightly larger value of $\overline{\omega}/{\gamma_G}$, and they seem have larger strain rate. On the contrary, sample number 7 and 8 have slightly smaller value of $\overline{\omega}/{\gamma_G}$ and they have smaller strain rate. If we take the rupture speed $V$ into account, \prettyref{eq:scale_link_morphology_energy} becomes
\begin{equation}
 \frac{R_G}{R_H} \sim \frac{\overline{\omega}(1 - \frac{V^2}{(B\beta)^2})}{\gamma_G} \frac{\Omega}{E_f},
\end{equation}
where $B$ is a constant of the order of 1 and $\beta$ is the shear-wave speed \citep{Kanamori2004}, because
\[G = G^*g(V) = \frac{\partial \Omega}{\partial s} (1 - \frac{V^2}{(B\beta)^2}).\]
But this doesn't explain the data and the rupture speeds $V$ aren't significantly different in those tests, so the rupture speeds $V$ doesn't have a significant effect here. One explanation is that for real materials, $\sigma^2_Y$ depends on the state of deformation and its history. Experiments discussed here show that high strain rate results in larger value of $\overline{\omega}$, i.e., on average, it needs more energy to remove or deform a protruding obstacle on the fracture surface during the friction. In the formation of the macroscopic crack and the protruding obstacles on the crack surface, high strain rate may make less plastic deformation inside the protruding obstacles, and hence more energy is needed to remove or deform them in the following deformation.

\section{Conclusions}

The dynamic energy balance is essential for earthquake studies. To interpret seismological data, crack models and sliding on a frictional surface (fault) models are widely used. From these two types of models, the macroscopically observable energy budget and the microscopic processes can be related through the fracture energy $G_c$.

The fault surface morphology is the direct result of the microscopic processes near the crack tip or on the frictional interface. Here we show that the dynamic energy balance in earthquakes can be expressed by fault surface morphology, and that they are quantitatively linked.

The direct shear experiments proves the predictions of the theoretical discussions, and show that the strain rate has crucial influence on the dynamic energy balance. The link predicted by the theoretical discussions has the property of material-independent, at least for dry brittle materials discussed here.

\section*{Acknowledgments}

This research was funded by the Fundamental Research Funds for the Central Universities (grant no. 2020QN29), the China Postdoctoral Science Foundation (Grant no. 2020M681759), and the State Key Program of National Natural Science Foundation of China (Grant no. 51934007).


\bibliography{elsarticle-template.bib}

\end{document}